# Two-dimensional Mott variable-range hopping transport in a disordered MoS$_2$ nanoflake

Jianhong Xue,[1] Shaoyun Huang,[1*] Ji-Yin Wang,[1] and H. Q. Xu[1,2+]

[1]*Beijing Key Laboratory of Quantum Devices, Key Laboratory for the Physics and Chemistry of Nanodevices and Department of Electronics, Peking University, Beijing 100871, China*

[2]*Division of Solid State Physics, Lund University, Box 118, S-221 00 Lund, Sweden*

**ABSTRACT**

The transport characteristics of a disordered MoS$_2$ nanoflake in the insulator regime are studied by electrical and magnetotransport measurements. The layered MoS$_2$ nanoflake is exfoliated from a bulk MoS$_2$ crystal and the conductance $G$ and magnetoresistance are measured in a four-probe setup over a wide range of temperatures. At high temperatures, we observe that $\log_{10} G$ exhibits a $-T^{-1}$ temperature dependence and the transport in the nanoflake dominantly arises from thermal activation. At low temperatures, where the transport in the nanoflake dominantly takes place via variable-range hopping (VRH) processes, we observe that $\log_{10} G$ exhibits a $-T^{-1/3}$ temperature dependence, an evidence for the two-dimensional (2D) Mott VRH transport. The measured low-field magnetoresistance of the nanoflake in the insulator regime exhibits a quadratic magnetic field dependence $\sim \alpha B^2$ with $\alpha \sim T^{-1}$, fully consistent with the 2D Mott VRH transport in the nanoflake.

Correspondence should be addressed to: +Professor H. Q. Xu (hqxu@pku.edu.cn) and *Dr. Shaoyun Huang (syhuang@pku.edu.cn)

Despite the rapid developments in $MoS_2$ layered materials and device applications, the nature of charge transport still remains elusive, since the experimentally measured carrier mobility in the materials is significantly lower than theoretical prediction[1-4]. Disorders, such as sulphur vacancies, charge impurities, surface absorbents, charged traps at $MoS_2$ layer-substrate interfaces[5-7], etc., strongly influence the transport properties of the layered $MoS_2$ materials. Generally, in a disordered system[8], transport can be divided into two different regimes, separated by the mobility edge $E_C$. Tuning the Fermi energy $E_F$ across the mobility edge causes a metal-insulator transition[9-11] from extended to localized states or vice versa. In the insulating regime, where carriers are all frozen to localized states at energies below the mobility edge $E_C$ at zero temperature, carrier transport can take place via thermal activation at relatively high temperatures and variable-range hopping (VRH) at low but finite temperatures. When the carrier transport is predominantly due to thermal activation, the conductance ($G$) exhibits a temperature ($T$) dependence of $\log_{10} G \sim -T^{-1}$ and, thus, the characteristics activation energy $E_a$ can be extracted from the Arrhenius plot of the measured conductance $G$ as a function of temperature $T$. When temperature goes higher than a critical value (~100 K in many compound semiconductor materials), phonon scattering can become significantly strong, leading to a characteristic temperature dependence that the conductance decreases with increasing temperature. At low temperatures, the current is carried via VRH processes in which a localized electron at the Fermi level moves to another localized state in an optimum hopping distance. The optimum distance is determined by the tradeoff between the lowest energy differences and the shortest hopping distances. In a noninteracting $d$-dimensional system, the density of states $N_F$ at the Fermi level is finite and the conductance can be well described by the Mott VRH mechanism[12] with $\log_{10} G \sim -T^{-p}$ where exponent $p=1/(d+1)$. Pollak[13-14], Efros and Shklovskii[15] pointed out that a soft gap can be opened up at the Fermi level by taking into account the long range electron-electron Coulomb interactions. The density of states at the Fermi level thus vanishes and the conductance should be described by the so called ES VRH mechanism[16-18] with exponent $p=1/2$ in $\log_{10} G \sim -T^{-p}$, independent of the dimension.

Due to the two-dimensional (2D) nature and the natural presence of disorders, layered $MoS_2$ offers a renewed platform to investigate VRH mechanisms of 2D transport. Currently, in layered $MoS_2$ systems, a $-T^{-1}$ dependence of $\log_{10} G$ is observed at relatively high temperatures and is attributed to thermally activated transport[19]. However, at relatively low temperatures in the experiments reported so far, both the Mott VRH and the ES VRH model can provide a satisfactory fitting to the measured conductance data and thus the transport mechanism cannot be uniquely determined in MoS$_2$ nanoflakes[20]. The two VRH mechanisms in a disordered MoS$_2$ system could also be distinguished by magnetotransport measurements[21]. In a common situation where the magnetoresistance shows a quadratic dependence on magnetic field as $\alpha B^2$, the coefficient $\alpha$ has different temperature dependences for the two VRH transport mechanisms. In the Mott VRH regime the coefficient $\alpha \propto T^{-3/(d+1)}$, whereas in the ES VRH regime the coefficient $\alpha \propto T^{-3/2}$. Thus, magnetotransport measurements could provide additional information to identify the transport mechanisms of VRH in disordered MoS$_2$ nanoflakes.

In this letter, we report on an experimental study of the transport characteristics of a disordered MoS$_2$ nanoflake by electrical and magnetotransport measurements in a four-probe setup over a temperature range of 6 to 300 K. The study is focused on the insulator regime where the Fermi energy $E_F$ lies below the mobility edge $E_C$. When $E_F$ is tuned close to $E_C$, the characteristics of thermally activated transport and phonon scattering are observed in the conductance measurements of the nanoflake. When $E_F$ is tuned far below $E_C$, the measured conductance in logarithmic scale shows a $-T^{-1}$ dependence at relatively high temperatures and a $-T^{-1/3}$ dependence at relatively low temperatures. A good quadratics magnetic field dependence $\sim \alpha B^2$ of the magnetoresistance is also observed in the nanoflake and a $T^{-1}$ dependence of coefficient $\alpha$ is extracted. These electrical and magnetotransport measurements provide for the first time a solid evidence that the Mott VRH rather than ES VRH transport is the dominant transport mechanism in a disordered MoS$_2$ nanoflake in the insulating regime at low temperatures.

The MoS$_2$ nanoflakes studied in this work are obtained by exfoliation from a commercially available bulk MoS$_2$ crystal. The exfoliated MoS$_2$ nanoflakes are transferred onto a highly doped silicon substrate covered with a 300-nm-thick layer of SiO$_2$ on top. Electrical contacts are prepared using electron-beam lithography (EBL) for pattern definition, electron-beam evaporation for deposition of 5-nm-thick titanium and 50-nm-thick gold, and lift-off process. Figure 1(a) shows an atomic force microscope (AFM) image of a fabricated device measured for this work and the schematic for the measurement circuit setup. The MoS$_2$ nanoflake in the device has a width of *W*~400 nm and a thickness of *t*~10 nm, see the AFM line scan measurements across an edge of the nanoflake shown in the lower panel of Figure 1(a). The device consists of four metal stripe contacts, which we labeled as contacts 1 to 4 as in Figure 1(a). These contacts have a width of 200 nm. The edge-to-edge distances between contacts 1 and 2, between contacts 2 and 3 and between contacts 3 and 4 are 80, 450 and 100 nm, respectively. The measurements are performed in a Physical Property Measurement System (PPMS) cryostat, which provides temperatures in a range of 300 to 2 K and magnetic fields up to 9 Tesla. The four-probe setup is adopted in the measurements, in order to exclude the contact resistances, in which a source-drain bias voltage $V_{ds}$ is applied between contacts 1 and 4, the channel current $I_{ds}$ and the voltage drop ($V_{23}$) between contacts 2 and 3 are simultaneously recorded, see the circuit setup in Figure 1(a). The Fermi level $E_F$ in the nanoflake is modulated by a back gate voltage $V_g$ applied to the silicon substrate.

Figure 1(b) shows the measured channel current $I_{ds}$ as a function of the voltage drop $V_{23}$ between contacts 2 and 3 for the device shown in Figure 1(a) at a fixed back gate voltage of $V_g = -30$ V and different temperatures. It is seen in the figure that the measured $I_{ds}$-$V_{23}$ curves are straight lines. The good linearity is found in all the measured $I_{ds}$-$V_{23}$ curves over a wide range of back gate voltages and of temperatures, which ensures that the transport characteristics of the MoS$_2$ channel are extracted from the measurements. Figure 1(c) shows the measured channel conductance $G = I_{ds}/V_{23}$ as a function of back gate voltage $V_g$ at different temperatures. It is seen that the device is a typical n-type transistor. The channel conductance shows very different temperature dependence at high back gate voltages (on the right side of point A) and at low back gate voltages (on the left side of point A), where the crossover point A is located at $V_g \sim -30$ V, as indicated by a black arrow in Figure 1(c). At the high back gate voltages (on the right side of point A), the conductance $G$ is increased with decreasing temperature and then becomes decreased with further lowering temperature. At the low back gate voltages (on the left side of point A), however, the conductance $G$ is monotonously decreased with decreasing temperature in the entire measured temperature range (from 300 to 6 K). The observed temperature dependence of the conductance at low temperatures indicates that the MoS$_2$ channel is in the insulating regime and the Fermi energy $E_F$ lies below the mobility edge $E_C$ throughout the entire measurement range of back gate voltages. The characteristic conductance increase with decreasing temperature observed at high back gate voltages and high temperatures arises from the interplay between the thermal activation transport and phonon scattering. This interplay phenomena could be better visualized by plotting the resistance $R_{23}$ as a function of temperature measured at different back gate voltages as shown in Figure 1(d). Here, we can clearly recognize that at a given high back gate voltage $V_g > -30$ V, there exists a characteristic temperature $T_M$, at which the resistance has a minimum. Apparently, $T_M$ increases with decreasing back gate voltage, see the yellow dashed line in Figure 1(d), and can reach a temperature as low as ~100 K at $V_g = 30$ V.

Physically, at such a high back gate voltage, the Fermi level $E_F$ is close to the mobility edge $E_C$ and a significant number of carriers can be excited to the extended states located at energies above the mobility edge at high temperatures. Thus, at $T > T_M$, the observed fact that the resistance decreases with decreasing temperature is mainly due to reduction in phonon scattering with decreasing temperature. However, at $T < T_M$, the phonon scattering becomes less important and the resistance becomes closely related to the number of carriers which are thermally excited to the extended states. As the back gate voltage $V_g$ decreases, the Fermi level $E_F$ is gradually moved away from the mobility edge $E_C$, leading to an increase in $T_M$ as seen in Figure 1(d).

Figure 2 shows the Arrhenius plot of the measured conductance as a function of temperature at different back gate voltages. In the temperature region of $80 \text{ K} < T < T_M$, i.e., the shaded part except for the upper-left corner region in the figure, relatively large thermal kinetic energy assisted transport dominates and the temperature dependence of the conductance can be well modelled by the thermally activated transport[19] as $G_a = G_0 e^{-E_a/k_B T}$. Here, $G_0$ is the conductance at the high temperature limit, $E_a = (E_C - E_F)$ is the activation energy, and $k_B$ is the Boltzmann constant. The extracted activation energy $E_a$ is shown in the inset of Fig. 2 as a function of the back gate voltage. The activation energy $E_a$ decreases linearly with increasing $V_g$ from -70 to -30 V and turns to saturate with further increasing $V_g$ to the positive side, in good agreement with the fact that the Fermi level moves closer to the mobility edge with increasing back gate voltage.

However, the thermally activated transport model does not describe the measurements in the low temperature region [the right, unshaded part of Figure 2] as seen from the deviations from the fitting lines in the region. Physically, in this low temperature region, VRH conduction becomes dominant and is responsible for the temperature dependent characteristics of the measured conductance. In theory, the conductance in VRH mechanisms can be described as

$$G = G_0 e^{-(\frac{T_p}{T})^p}, \tag{1}$$

where $T_p$ is the characteristic temperature and the exponent $p$ depends on VRH mechanism. For the Mott VRH conduction, the exponent $p$ is dimension dependent and has a value of $p=1/3$ in a 2D system. For the ES VRH conduction, the exponent is dimension independent and has a value of $p=1/2$. To identify whether the Mott VRH mechanism or the ES VRH mechanism play a dominant role in determining the transport characteristics of the MoS$_2$ nanoflake, we plot the measured conductance as a function of $T^{-1/2}$ and of $T^{-1/3}$ in Figures 3(a) and (b), respectively. It is seen in Figure 3(a) that the measured data cannot be fitted by straight lines, indicating that the ES VRH mechanism does not describe the transport behavior of the nanoflake in this low temperature range. However, in Figure 3(b), straight-line fits agree excellently with the measured data over the entire temperature range of 6 to 80 K. Thus, the transport in the MoS$_2$ nanoflake in this temperature range is most likely governed by the 2D Mott VRH process. We have also checked and fitted our data against the 3D Mott VRH model (see Supporting Information), in which the exponent takes a value of $p=1/4$, and found large deviations, indicating that the transport in the nanoflake should be of the 2D nature. The 2D nature of the transport in our nanoflake is also supported by the angular dependent magnetoresistance measurements as shown in Supporting Information.

Employing the 2D Mott VRH mechanism, we can further analyze the temperature dependence of the conductance to extract the characteristic temperature $T_{1/3}$ and localization length $\xi_{loc}$ in the nanoflake at different gate voltages $V_g$. The results are shown in the inset of Figure 3(b). It is seen that $T_{1/3}$ monotonically decreases from ~$10^4$ to ~$10^2$ K with increasing $V_g$ from -30 to 5 V. Theoretically, $T_{1/3}$ is related to the localization length as $T_{1/3} = 13.8/(k_B N_F \xi_{loc}^2)$, where $N_F$ is the density of states at Fermi level. The $N_F$ can be determined from the gate voltage dependence of the activation energy $E_a$ by taking into account the quantum capacitance $C_d$ as[11,22]

$$\frac{dE_F}{dV_g} = -\frac{dE_a}{dV_g} = \frac{eC_g}{C_g + C_d}, \qquad (2)$$

with $C_d = e^2 N_F$. Here, $C_g$ is the unit area capacitance and can be obtained from $C_g = \frac{\varepsilon_0 \varepsilon}{d}$ with $\varepsilon_0$ being the vacuum permittivity, $\varepsilon$ and $d$ being the dielectric constant and layer thickness of SiO$_2$. The calculated values of $C_d$ and $N_F$ are 8 $\mu$F·cm$^{-2}$ and $5 \times 10^{13}$ eV$^{-1}$·cm$^{-2}$, respectively. Thus, the derived localization length in the nanoflake is found to increase from 5.6 to 48.5 nm with increasing $V_g$ from -30 to 5 V, as shown in the inset of Fig. 3(b).

Figures 4(a) show the magnetoresistance characteristics at different temperatures in the 2D VRH transport regime at $V_g = -20$ V with the magnetic field $B$ applied perpendicular to the MoS$_2$ nanoflake. Here, the MoS$_2$ nanoflake is in the *x-y* plane, the current flow is along the *x* axial direction, and the magnetoresistance is defined as $\Delta R_{23} = [R(B) - R(B=0)]/R(B=0)$. Clearly, the magnetoresistance shows a positive quadratic dependence on magnetic field $B$ at temperatures below ~50 K. In the wave-function shrinkage model[23-26], the positive quadratic magnetoresistance is attributed to the contraction of the electronic wave function at traps in a magnetic field, thus leading to a reduction of hopping probability. Quantitatively, in the wave-function shrinkage model, the magnetoresistance in the Mott VRH regime can be expressed as[21]

$$\Delta R_{23} = 0.0893 \left[\frac{e^2 \xi_{loc}^4}{36\hbar^2}\right]\left(\frac{T_{Mott}}{T}\right) B^2 = \alpha^{Mott} \cdot B^2, \quad (3)$$

where $T_{Mott}$ is a characteristic temperature, $\hbar$ is the reduced Planck constant and the prefactor $\alpha^{Mott}$ is proportional to $T^{-1}$. Note that in the ES VRH regime, the magnetoresistance can be found by replacing $(T_{Mott}/T)$ in Equation (3) with $(T_{ES}/T)^{3/2}$. Thus, the magnetoresistance also shows a positive quadratic magnetic field dependence, but the prefactor $\alpha^{ES}$ is proportional to $T^{-3/2}$ instead of $T^{-1}$. We fit the positive quadratic magnetoresistance curves at different temperatures in Fig. 4(a) based on the wave-function shrinkage models in both the Mott VRH and the ES VRH regime. The derived prefactors $\alpha^{Mott}$ and $\alpha^{ES}$ are shown in Fig. 4(b). Clearly, the extracted $\alpha^{ES}$ as a function of $T^{-3/2}$ shown in the inset of Fig. 4(b) exhibits a significant deviation, while the extracted $\alpha^{Mott}$ displays an excellent agreement with the $T^{-1}$ dependence. Thus, the 2D Mott VRH mechanism rather than the ES VRH mechanism is identified for the electron transport in the $MoS_2$ nanoflake at low temperatures, in good agreement with the analysis shown in Fig. 3.

In conclusion, the transport characteristics of a disordered $MoS_2$ nanoflake have been investigated in details over a wide range of temperatures in the insulator regime, where the Fermi level $E_F$ in the nanoflake is tuned with use of the back gate voltage to lie below the mobility edge $E_C$. At relatively high temperatures, the nanoflake exhibits activation transport characteristics. The activation energy $E_a = E_C - E_F$, which measures the energy distance between the mobility edge $E_C$ and the Fermi energy $E_F$, is extracted in the nanoflake. It is found that the activation energy $E_a$ decreases with increasing back gate voltages at low back gate voltages and turn to saturate towards zero at high back gate voltages. At sufficiently low temperatures, the transport characteristics of the nanoflake are found to be governed by VRH processes. To identify whether the Mott or the ES VRH mechanism plays a dominant role in system at this low temperature region, the temperature dependent conductance and magnetoresistance have been measured and analyzed. It is found that in this low temperature region the conductance in logarithmic scale shows a $-T^{-1/3}$ temperature dependence and the prefactor in the quadratic magnetic field dependent magnetoresistance scales with temperature as $T^{-1}$. These results provide exclusive evidences that the 2D Mott VRH transport is the dominant transport mechanism at low temperatures in the insulating regime of the disordered $MoS_2$ nanoflake.

**ACKNOWLEDGEMENTS**

This work is supported by the Ministry of Science and Technology of China through the National Key Research and Development Program of China (Grant Nos. 2017YFA0303304, 2016YFA0300601, 2017YFA0204901, and 2016YFA0300802), and the National Natural Science Foundation of China (Grant Nos. 91221202, 91421303, and 11274021). HQX also acknowledges financial support from the Swedish Research Council (VR).

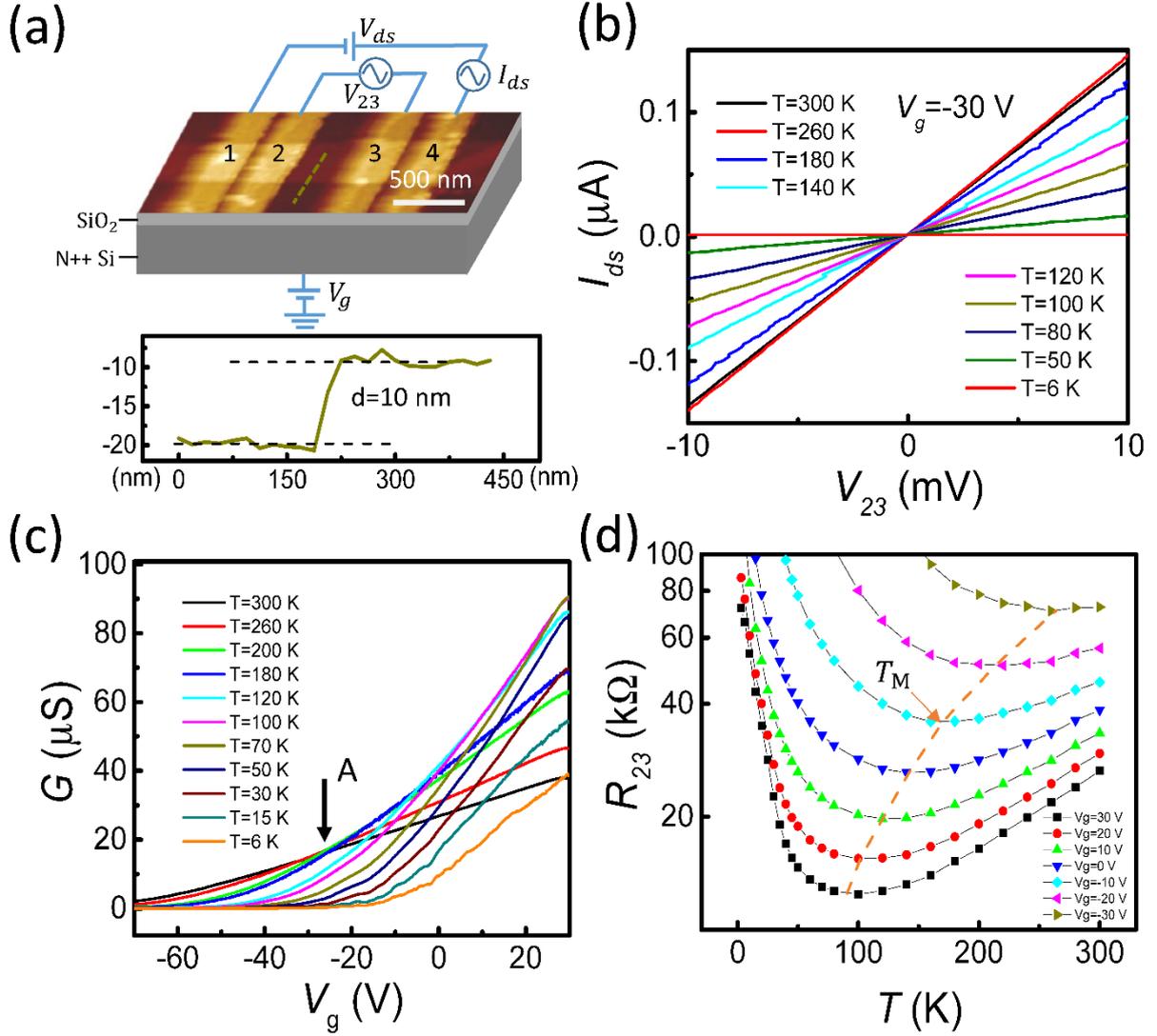

Figure 1. (a) AFM image of the MoS$_2$ nanoflake device and schematic view of the device layer structure and measurement setup (top panel), and height profile measured using AFM along the dashed line in the AFM image (bottom panel). Here it is shown that the MoS$_2$ nanoflake in the device has a thickness of 10 nm. (b) Source-drain current $I_{ds}$ vs. voltage $V_{23}$ measured for the device at temperatures from 6 to 300 K and at back gate voltage $V_g = -30$ V. (c) Transfer characteristics of the field-effect device at temperatures from 6 to 300 K. (d) Resistance of the nanoflake plotted against temperature $T$ at different back gate voltages. $T_M$ marks the temperature position at which the resistance has a minimum in the curve measured at each back gate voltage. The dashed line connecting the values of $T_M$ at different back gate voltages is the guide to the eyes.

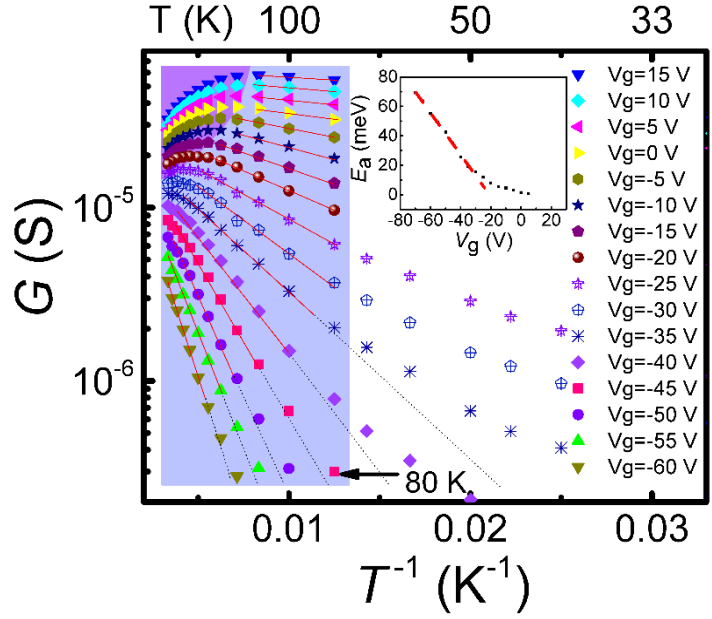

Figure 2. Conductance *G* in logarithmic scale (Arrhenius plot) plotted against $T^{-1}$ (inverse of temperature) for the device at different back gate voltages. The region on the left marked by grey color is for the measurements at temperatures higher than ~80 K, at which the transport in the nanoflake is well described by the thermal activation mechanism. The upper-left corner marked by light purple color is the region of the measurements at high temperatures and high positive gate voltages, where the characteristics of phonon scattering in the layered $MoS_2$ is observed. Lines in the grey colored region are straight line fits to the measured data. The inset shows the extracted activation energy $E_a$ from the straight line fits in the grey color region as a function of back gate voltage $V_g$.

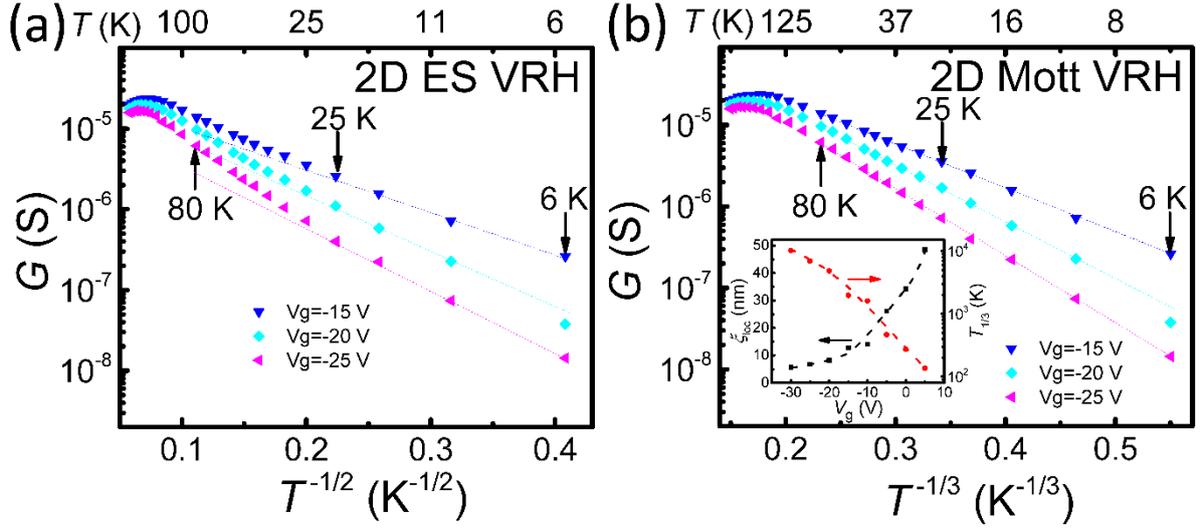

Figure 3. (a) Conductance $G$ in logarithmic scale plotted against $T^{-1/2}$ for the device measured at different back gate voltages. Lines are straight line (2D ES VRH theory) fits to the measurement data at low temperatures. Clearly, the straight line fits do not describe the measurement data at temperatures of 25 to 80 K. (b) The same as in (a) but plotted against $T^{-1/3}$. Lines are straight line (2D Mott VRH theory) fits to the measurement data at temperatures of 6 to 80 K. Here, excellent fits are obtained over this range of temperatures. The inset shows the extracted characteristic temperature $T_{1/3}$ and localization length $\xi_{loc}$ as a function of back gate voltage $V_g$ for the $MoS_2$ nanoflake based on the 2D Mott VRH theory.

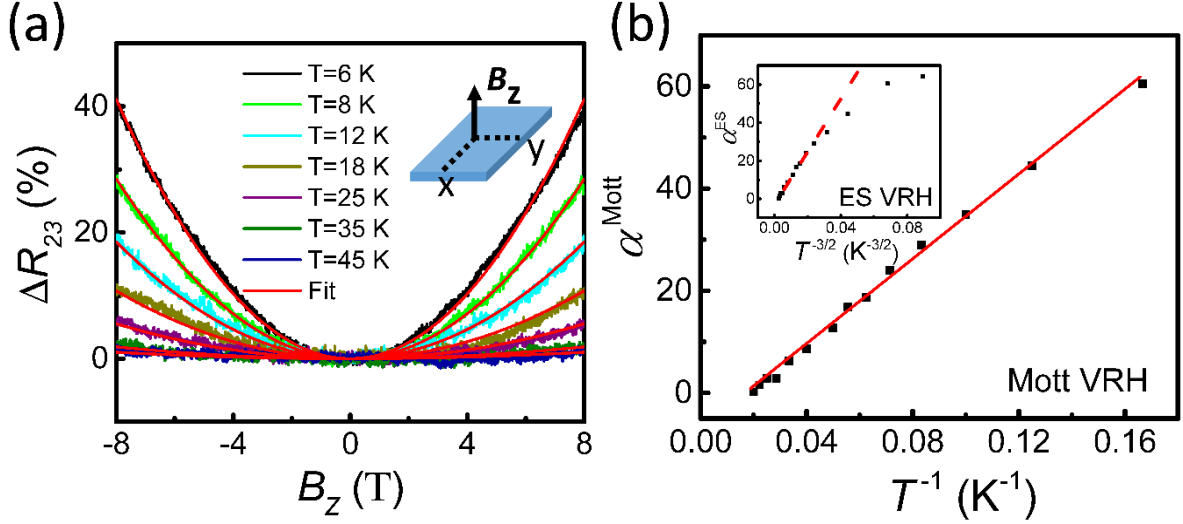

Figure 4. (a) Magnetoresistance as a function of magnetic field $B_z$ applied perpendicular to the MoS$_2$ nanoflake plane (as shown in the inset) measured at back gate voltage $V_g = -20$ V and at different temperatures. Red solid lines are fits to the measured data based on the wave-function shrinkage model. (b) Prefactor $\alpha^{Mott}$ as a function of $T^{-1}$ extracted from the measured magnetoresistance curves at different temperatures $T$. The red solid line presents the predicted values of $\alpha^{Mott}$ by the wave-function shrinkage model in the Mott VRH regime. The inset shows the prefactor $\alpha^{ES}$ extracted from the same magnetoresistance measurements as a function of $T^{-3/2}$. The dashed line in the inset shows the results that would be predicted by the wave-function shrinkage model in the ES VRH regime.

# Supplemental Materials for
# Observation of two-dimensional Mott variable-range hopping transport in a disordered MoS$_2$ nanoflake

Jianhong Xue,[1] Shaoyun Huang,[1,*] Ji-Yin Wang,[1] and H. Q. Xu[1,2,+]

[1]*Beijing Key Laboratory of Quantum Devices, Key Laboratory for the Physics and Chemistry of Nanodevices and Department of Electronics, Peking University, Beijing 100871, China*

[2]*Division of Solid State Physics, Lund University, Box 118, S-221 00 Lund, Sweden*

Correspondence should be addressed to: +Professor H. Q. Xu (hqxu@pku.edu.cn) and *Dr. Shaoyun Huang (syhuang@pku.edu.cn)

## 1. Temperature dependence of the electron mobility in the MoS$_2$ nanoflake

Figure 1(c) of the main article shows the measured channel conductance $G=I_{ds}/V_{23}$ of the studied MoS$_2$ nanoflake device as a function of back gate voltage $V_g$ at different temperatures, see Figure 1(a) in the main article for the device structure and the measurement setup. The electron mobility of the nanoflake at different temperatures can be extracted from the measured conductance curves shown in Figure 1(c) of the main article using the following equation

$$\mu = \frac{L_{23}}{W \times C_g} \times \frac{dG}{dV_g}, \quad (1)$$

where the channel length $L_{23}$ is 650 nm and the channel width $W$ is 400 nm. $C_g$ is the unit area capacitance, which can be estimated using $C_g = \frac{\varepsilon_0 \varepsilon}{d}$, where $\varepsilon_0$ is the vacuum permittivity, $\varepsilon = 3.9$ is the dielectric constant of SiO$_2$, $d = 300$ nm is the thickness of SiO$_2$. Figure 1 in this Supporting Information shows the extracted mobility from the measured conductance shown in Figure 1(c) of the main article. The results shown in Figure 1 manifest distinctly different temperature dependences at low and at high temperatures. In the low temperature region (the left side of the figure), the mobility decreases with decreasing temperature, which is consistent with the fact that the nanoflake is in the insulating regime at all the gate voltages considered in this work. In the high temperature region (the right side of the figure), the mobility decreases with increasing temperature, showing the characteristic influence of phonon scattering on the electron transport in the nanoflake. As discussed in the main article, in this high temperature region, the transport is predominantly carried out by the carriers which are thermally excited to the extended states located above the mobility edge. However, at these high temperatures, phonons become active and phonon scattering plays a dominant role in limiting carrier transport, resulting in a characteristic dependence of the carrier mobility on temperature, $\mu \propto T^{-\gamma}$, where constant $\gamma$ depends on the specific phonon scattering mechanisms. The fitting at the high temperature side of Figure 1(a) gives a value of $\gamma \approx 1.74$, which is much close to the theoretically predicted value for MoS$_2$ optical phonon scattering in monolayer MoS$_2$ ($\gamma \approx$ 1.52)[1] and is very different from that in bulk crystals ($\gamma \approx 2.6$)[2]. The experimental results are consistent with the fact that the thickness of the MoS$_2$ nanoflake is 10 nm in the device.

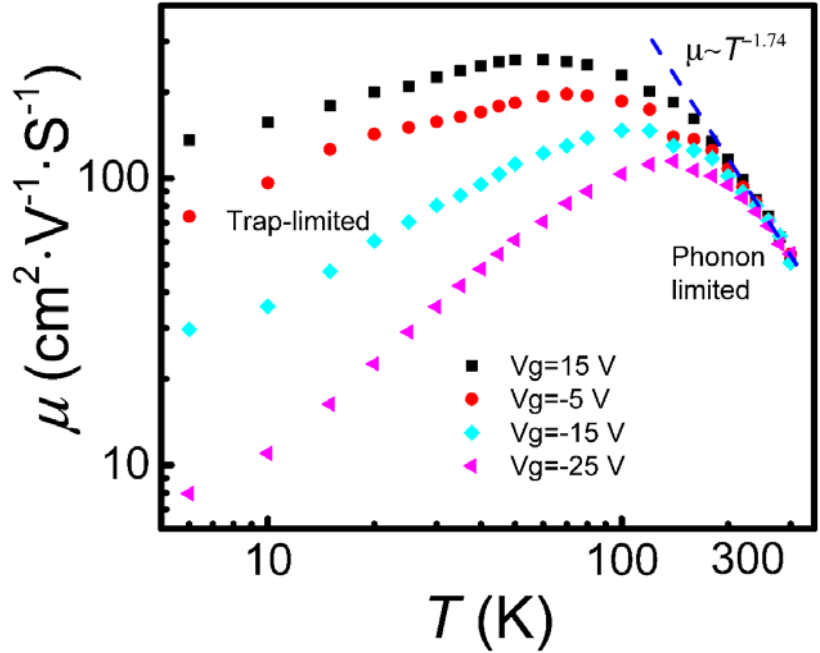

Figure 1. Electron mobility $\mu$ extracted for the MoS$_2$ nanoflake studied in the main article as a function of temperature at different gate voltages. At low temperatures, $\mu$ decreases with decreasing temperature, showing that the nanoflake is in the insulating regime at the gate voltages considered. At high temperatures, $\mu$ decreases with increasing temperature as $\mu \propto T^{-1.74}$, implying that the carrier transport is dominantly limited by optical phonon scattering in the nanoflake.

**2. Fitting our measurement data to the 3D Mott VRH model**

In the main article, we show that the measured conductance at low temperatures of 6 K < T < 80 K are fitted excellently by the 2D Mott variable-range hopping (VRH) model, but cannot be well described by the 2D ES VRH model. Here we show in Figure 2 that the measurement data could not be fitted by the 3D Mott VRH model. In the 3D Mott VRH model, the conductance in logarithmic scale is related to temperature as $-T^{-1/4}$. In Figure 2, we replot the measurements data (in logarithmic scale) as a function of $T^{-1/4}$. It is clearly seen that a straight line fit in the high temperature region from 80 to 25 K does not extended to the low temperature region from 25 to 6 K well. Thus, the 3D Mott VRH model fails to describe the transport behavior in the nanoflake in the full temperature range from 80 to 6 K.

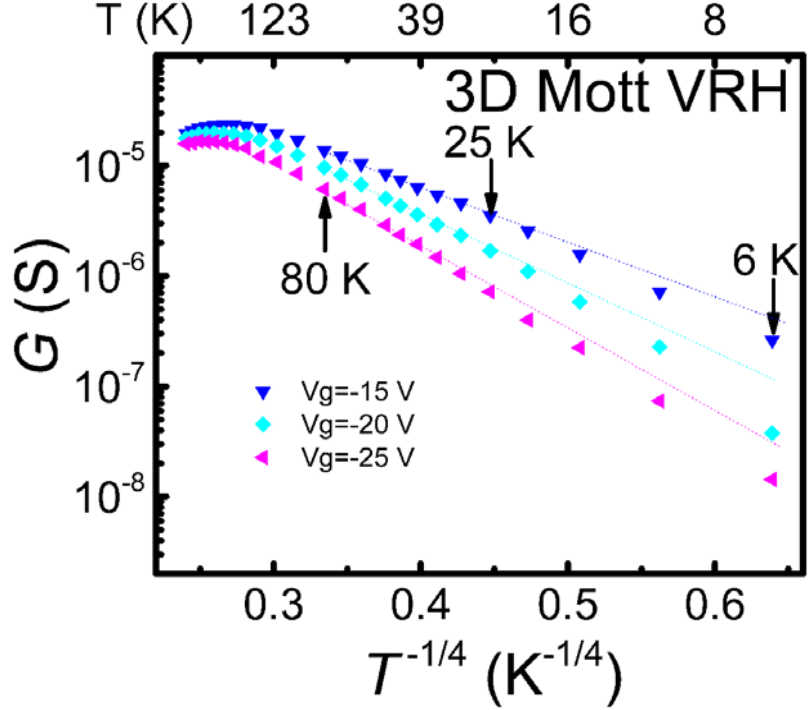

Figure 2. Measured conductance $G$ (in logarithmic scale) plotted against $T^{-1/4}$ at three representative back gate voltages. Clearly, the data in the temperature range of 6 to 80 K could not be fitted using single straight lines.

**3. Magnetoresistance measurements with the magnetic fields applied at different angles**

To show further that the transport in the $MoS_2$ nanoflake is predominantly of 2D nature, we measure the magnetoresistance of the nanoflake device with the magnetic fields applied at different angles $\theta$. The results of the measurements for the device at temperature $T=6K$ and back gate voltage $V_g=-20$ V are shown in Figure 3(a). Here, as shown in the inset of Figure 3(a), the nanoflake is in the $x$-$y$ plane, the transport takes place along the $x$ direction, and the magnetic fields are applied in the $y$-$z$ plane, i.e., always perpendicular to the current direction. The magnetoresistance is defined as $\Delta R_{23} = [R(B) - R(B = 0)]/R(B = 0)$, where $R = V_{23}/I_{ds}$, see Figure 1(a) of the main article for the measurement circuit setup. At $\theta = 90°$, the magnetoresistance shows a positive quadratic magnetoresistance as we discussed in the main article. At $\theta = 0°$, a weak negative magnetoresistance is observed. This weak negative magnetoresistance arises from the finite thickness (~10 nm) of the nanoflake and thus the suppression of back scattering by the top and bottom surfaces of the nanoflake by the in-plane magnetic field. However, the thickness of the nanoflake is very small, the negative magnetoresistance is small in magnitude, e.g., it is less than 3% even at the in-plane magnetic field of 8 T. Figure 3(b) shows the normalized magnetoresistance, $\Delta R_{23}^+ = \Delta R_{23} - \Delta R_{23}^{\theta=0°} \cdot \cos(\theta)$, as a function of perpendicular component of the magnetic field $B_z = B \cdot \sin\theta$. Here, the $\Delta R_{23}^{\theta=0°}$ is the measured magnetoresistance at in-plane magnetic fields and can be acquired from Figure 3(a) at $\theta = 0°$. It is seen that the normalized magnetoresistance is solely dependent on the perpendicular component of the applied magnetic field. Overall, the results shown in this supplementary note imply that the transport in the nanoflake is predominantly of the 2D nature, but a small 3D transport characteristic could be present which may cause some small but observable deviations from the predictions of 2D transport theory.

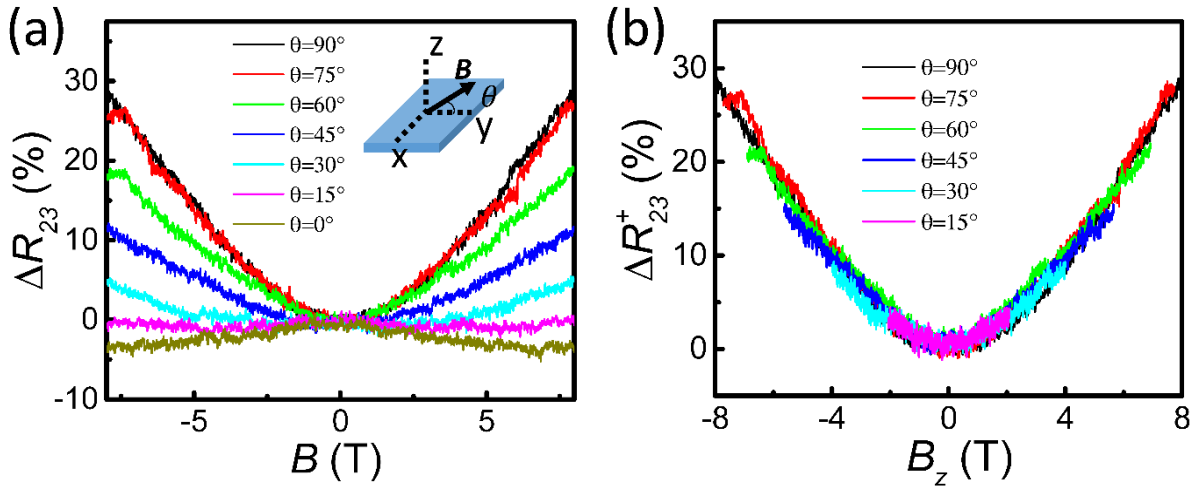

Figure 3. (a) Magnetoresistance measured for the nanoflake device with the magnetic fields applied at different orientations $\theta$ as shown in the inset at temperature $T$= 6 K and at back gate voltage $V_g$= −20 V. A positive quadratic magnetoresistance is observed at $\theta = 90°$ and a very weak magnetoresistance is observable at $\theta = 0°$. (b) Normalized magnetoresistance plotted against the perpendicular component of the applied magnetic field $B_z$. Here, the normalized magnetoresistance is obtained from (a) by subtracting the in-plane contributions from the measured magnetoresistance values.